\documentclass[9pt, conference]{IEEEtran}
\IEEEoverridecommandlockouts
\usepackage{cite}
\usepackage{amsmath,amssymb,amsfonts}
\usepackage{algorithmic}
\usepackage{graphicx}
\usepackage{textcomp}
\usepackage{xcolor}
\usepackage{blindtext}
\usepackage{hyperref}
\usepackage{pgfplots}
\usepackage{xcolor}
\usepackage{booktabs} 
\usepackage{bbm}
\usepackage{bm}
\usepackage{multirow}
\usepackage{siunitx}
\def\BibTeX{{\rm B\kern-.05em{\sc i\kern-.025em b}\kern-.08em
    T\kern-.1667em\lower.7ex\hbox{E}\kern-.125emX}}
\begin{document}

\title{ASTAR-NTU solution to AudioMOS Challenge 2025 Track1}

% \author{\IEEEauthorblockN{1\textsuperscript{st} Anonymous Author}
% \IEEEauthorblockA{
% \and
% \IEEEauthorblockN{2\textsuperscript{nd} Anonymous Author}
% \IEEEauthorblockA{
% \and
% \IEEEauthorblockN{3\textsuperscript{rd} Anonymous Author}
% \IEEEauthorblockA{
% \and
% \IEEEauthorblockN{4\textsuperscript{th} Anonymous Author}
% \IEEEauthorblockA{
% \and
% \IEEEauthorblockN{5\textsuperscript{th} Anonymous Author}
% \IEEEauthorblockA{
% \and
% \IEEEauthorblockN{6\textsuperscript{th} Anonymous Author}
% \IEEEauthorblockA{

\author{
Fabian Ritter-Gutierrez$^{1\dagger}$, Yi-Cheng Lin$^{2\dagger}$, Jui-Chiang Wei$^{2\dagger}$, 
Jeremy H.M. Wong$^{1}$, Nancy F. Chen$^{1}$, Hung-yi Lee$^{2}$
\\
\textit{$^1$Institute for Infocomm Research (A$^*$STAR)},  \textit{$^2$National Taiwan University}, \textit{$^\dagger$  Same Contribution} \\

\texttt{stufarg@i2r.a-star.edu.sg}\\
}
\maketitle

\begin{abstract}
Evaluation of text-to-music systems is constrained by the cost and availability of collecting experts for assessment. AudioMOS 2025 Challenge track 1 is created to automatically predict music impression (MI) as well as text alignment (TA) between the prompt and the generated musical piece. This paper reports our winning system, which uses a dual-branch architecture with pre-trained MuQ and RoBERTa models as audio and text encoders. A cross-attention mechanism fuses the audio and text representations. For training, we reframe the MI and TA prediction as a classification task. To incorporate the ordinal nature of MOS scores, one-hot labels are converted to a soft distribution using a Gaussian kernel. On the official test set, a single model trained with this method achieves a system-level Spearman's Rank Correlation Coefficient (SRCC) of 0.991 for MI and 0.952 for TA, corresponding to a relative improvement of 21.21\% in MI SRCC and 31.47\% in TA SRCC over the challenge baseline.

%Evaluation of text-to-music (TTM) systems is constrained by the cost and availability of collecting experts for music quality assessment. As such, AudioMOS 2025 Challenge track 1 is created to automatically predict music impression (MI) as well as text alignment (TA) between the prompt and the generated musical piece. This paper reports our winning system, which uses a dual-branch architecture with pre-trained MuQ and RoBERTa models as audio and text encoders. A cross-attention mechanism fuses the audio and text representations. For training, we reframe the MI and TA prediction as an original classification task. To incorporate the ordinal nature of MOS scores, one-hot target labels are converted to a soft distribution using a Gaussian kernel. On the official test set, a single model trained with this method achieves a system-level Spearman's Rank Correlation Coefficient (SRCC) of 0.991 for MI and 0.952 for TA. These results correspond to a relative improvement of 21.21\% in MI SRCC and 31.47\% in TA SRCC over the challenge baseline.

\end{abstract}
\section{Introduction}

Recent advances in generative modeling have enabled text‐to‐music (TTM) systems to produce high‐fidelity audio from natural‐language prompts \cite{audioldm,audioldm2,AudioXDT}. Despite these gains, evaluating musical quality remains a subjective, expert‐driven task where musicologists assign Mean Opinion Scores (MOS) on a 1–5 scale to judge aspects such as rhythm, harmony, among others. Recruiting and coordinating experts for MOS annotation is costly and slow, creating a bottleneck as TTM model development accelerates.

To overcome this limitation, we need automatic, scalable estimators of human judgments that jointly capture (1) intrinsic audio quality (artifact, musicality, and dynamic range), and (2) alignment between generated music and its prompt. The AudioMOS 2025 Challenge Track 1, based on the MusicEval dataset \cite{musiceval}, provides paired MOS annotations for Music Impression (MI) and Text Alignment (TA).

In this paper, we introduce DORA-MOS (Distributional Ordinal Ranking for Audio MOS), which uses a dual‐branch architecture that leverages pre‐trained audio and text encoders, applies cross‐modal attention to align audio/text representations, and reframes MOS prediction as an ordinal-aware classification task by softening one-hot labels with a Gaussian kernel to better optimize rank‐based evaluation metrics. Our method reached the 1st place in the AudioMOS Challenge 2025 track 1. 
\section{Proposed Method}
\vspace{-1mm}
\subsection{Data Analysis / Data Partition}

We use the MusicEval dataset \cite{musiceval}, which contains outputs from 31 TTM systems \cite{musiclm, musicgen, audioldm, audioldm2, audiogen, make_an_audio, noise2music, musecoco, ernie, mousai} on 384 fixed prompts, each rated by five experts for Music Impression (MI) and Text Alignment (TA). The original dev set did not have data for SystemIDs 06 and 18. To ensure evaluations across all systems and to evaluate the generalization of our models to unseen data, we stratify the provided training set and split it by system ID and score distributions. The resulting stratified dataset has a similar system ID and score distribution to the original train set, but the created dev set has a slightly more uniform distribution across system IDs. Any hyperparameter optimization was done in the stratified dataset, reserving the original dev set solely for final unseen data evaluation. Since the original test set contains no unseen models and musical genres for the text prompts, we treat this task as in-domain and forego external data or augmentation techniques.

\begin{figure}[t!]
    \centering
    \includegraphics[width=\columnwidth]{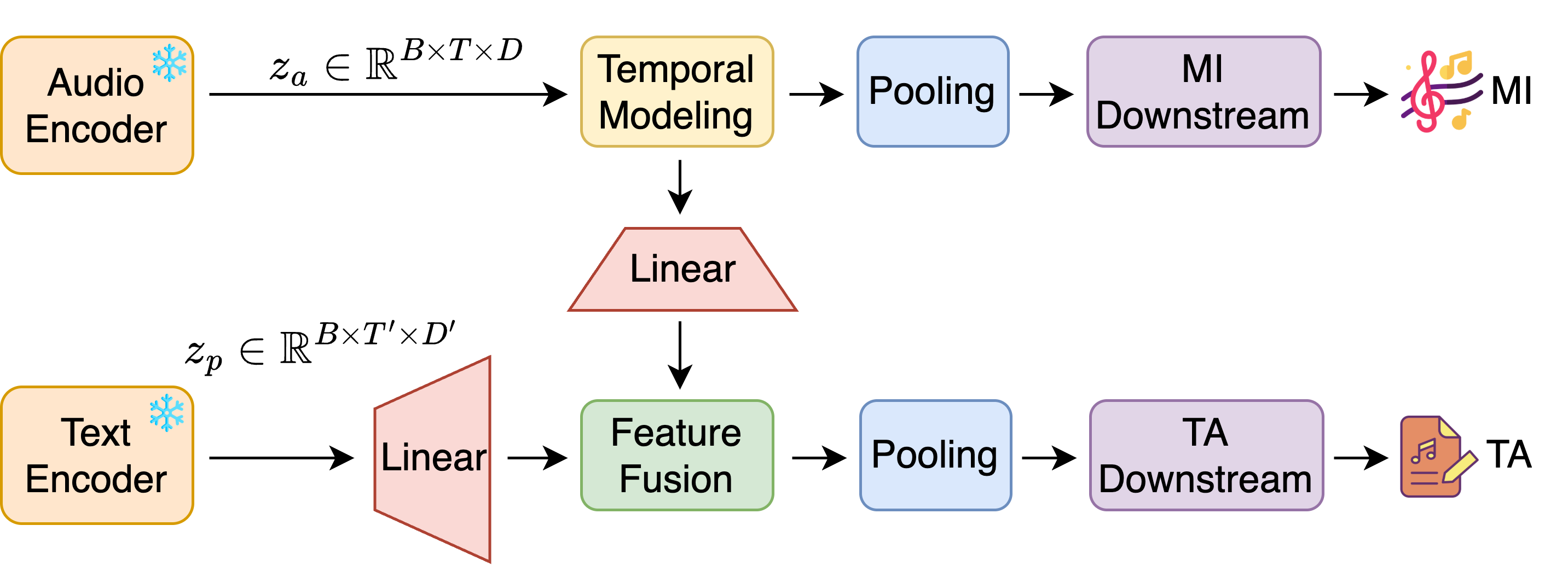} 
    \caption{Proposed dual-branch architecture for MI and TA prediction}
    \label{fig:main_figure}
\end{figure}
\subsection{Model Architecture}\label{subsec:model_architecture}

Pre-trained audio‐text models such as CLAP \cite{clap} and MuQ-Mulan \cite{muq} learn joint representations optimized for semantic similarity between an audio clip and its prompt. However, predicting the text alignment (TA) score may need a different audio/text alignment to model the ordinal assessment of prompt–audio coherence. To explicitly learn this alignment, we introduce a dual‐branch architecture (Fig.~\ref{fig:main_figure}) that leverages frozen unimodal encoders and a task‐specific feature fusion strategy that predicts TA scores.

% Models such as CLAP \cite{clap} or MuQ-Mulan \cite{muq} have been pre-trained to learn learning joint representations of audio and text. CLAP is the model used in the baseline; we add results in Table \ref{tab:evaluation_results_final_revised} by changing CLAP to MuQ-Mulan. The hypothesis in our proposed system is that such joint audio/text representations are not suitable for the challenge task. The internal alignment learned by CLAP or Muq-Mulan is geared towards semantic similarity between an audio and a text prompt. However, the TA prediction score requires determining how well, on an ordinal scale, a given audio sample aligns with its prompt. This specific task may require a different more specialized form of alignment more subtle to the features that influence human perception of quality and coherence. For this reason, we propose a model, depicted in Figure \ref{fig:main_figure}, designed to predict MI and TA scores through a dual-branch architecture that processes features from frozen, pre-trained encoders. The core of our approach is to leverage strong unimodal representations and learn a task-specific fusion strategy between the modalities.

The proposed architecture involves an audio encoder, a text encoder, a downstream model branch to model MI score, and a downstream model branch to model TA score. In order to model the TA score, we use cross-attention feature fusion that learns the correspondence between the input text prompt, the audio embeddings, and their respective coherence TA score.

\textbf{Feature Extraction:} Let $X_a$ be the input audio and $X_p$ its corresponding text prompt. We use MuQ \cite{muq} as the frozen audio encoder which process $X_a$, we take MuQ last layer representation $z_a \in \mathbb{R}^{B,T,D}$, where $B$ represent the batch size, $T$ the sequence dimension and $D$ the feature dimension. Similarly, the text encoder is composed of the RoBERTA model \cite{roberta}, which receives $X_p$ and outputs an embedding $z_p \in \mathbb{R}^{B,T',D'}$. We also evaluated CLAP, MERT, and MuQ-Mulan for audio feature extraction, but MuQ delivered the strongest performance.

\textbf{Music Impression (MI) Branch:} 
% The MI branch is designed to assess the intrinsic quality of the music and operates solely on the audio modality. It first, performs temporal modelling of the Muq features $z_a$ to capture long-range dependencies. To do so, a single layer of transformer is used with 4 self-attention heads. The output is a context-aware audio sequence $z'_a \in \mathbb{R}^{B,T,D}$.
% After obtaining the context-aware representations $z'_a $, we aggregate them into a single-sided vector $v_{MI}$ using attention-pooling which learns to weight the importance of different time steps. This vector is fed to the Music Quality Downstream Model, which consists of 2 multilayer perceptron (MLP) layers with ReLU activations. We finally predict a distribution over bins instead of a single scalar value, which will be explained further in the following section. 
The MI branch is designed to assess the intrinsic quality of the music and operates solely on the audio modality. A single transformer layer with four self-attention heads captures long-range dependencies in music features \(z_{a}\), producing context-aware embeddings. These embeddings are aggregated by attention pooling into a fixed-length vector, which is then passed through a two-layer multilayer perceptron (MLP) with ReLU activations to predict a MI score. We predict a distribution over 20 score bins rather than scalar values.

\textbf{Text Alignment (TA) Branch:} 
The TA branch is designed to evaluate the alignment between the audio and the text prompt. Two out of the three best-performing models follow the principle shown in Fig. \ref{fig:main_figure}, where we first project both the context-aware sequence and text embeddings to a common feature dimension before the cross-attention feature fusion. An additional system, called \textbf{DORA-MOS (Decoupled)}, trains the TA score by cross-attending the projected MuQ $z_a$ features with the projected RoBERTA text embeddings. Such an approach ensures that backpropagation on the temporal modelling block occurs not from TA gradients, but rather only from MI gradients. Regardless of the audio input, the cross-attention uses the text features as the query and the audio component as the key and value, learning text-informed audio representations. Finally, similarly to the MI branch, the text-informed representations are aggregated with attention-pooling %This vector, which encapsulates the learned audio-text alignment, is 
and then passed through a separate 2-layer MLP with ReLU activations to predict TA bins.

%This way, the model learn attention scores for the audio sequence that aligns with the input text prompt. 
%The output is then a text-informed audio representation $z'_{fused}$. 

\subsection{Training on an ordinal distribution rather than scalar values}

% The baseline approach for this task trains the model using an L1 loss, directly regressing the model's output to the ground-truth MOS score. This approach may not be optimal for a task where the primary evaluation metric is Spearman's Rank Correlation Coefficient (SRCC). SRCC measures the monotonic relationship between two variables, meaning it is concerned with preserving the rank order of predictions, not their precise values. A model can achieve a high SRCC even if its absolute predictions are systematically shifted, as long as it correctly identifies which audio clips are better or worse than others.

% Motivated by this, we reframe the task from regression to classification. This shift in perspective aligns more directly with optimizing rank order. We discretize the continuous MOS scale of 1 to 5 into 20 uniform bins. A model trained to classify an audio clip into the correct bin implicitly learns to place it in the correct rank relative to others.

% However, a naive classification approach using a standard cross-entropy loss with one-hot encoded targets is fundamentally flawed for an ordinal scale. Such a loss function treats all incorrect bins as equally wrong. For instance, if the actual score falls into bin N, a prediction for the adjacent bin (N+1) receives the same penalty as a prediction for the distant bin (N+10). This ignores the inherent ordinal structure of MOS ratings, where a small prediction error should be penalized far less than a large one.

The baseline approach for this task trains the model using an L1 loss over scalar values. Standard L1 regression minimizes mean absolute error but does not directly optimize ranking metrics such as SRCC. SRCC measures the monotonic relationship between two variables, meaning it is concerned with preserving the rank order of predictions, not their precise values. A model can achieve a high SRCC even if its absolute predictions are shifted, as long as it correctly identifies which audio clips are better or worse than others. To better align training with SRCC, we convert the MOS prediction into a classification task over $K=20$ equal-width bins spanning 1 to 5. A model trained to classify an audio clip into the correct bin is implicitly learning to place it in the correct rank position relative to others. However, a standard cross‐entropy loss with one‐hot targets ignores the ordinal structure of MOS. Misplacing a sample by one bin (N→N+1) incurs the same penalty as misplacing it by ten bins (N→N+10), even though larger errors should be penalized more heavily.

To address this limitation and introduce ordinality into the classification task, we propose a \textbf{Gaussian label softening} technique. Instead of a hard one-hot vector, we generate a soft target distribution for each training sample. The distribution is defined by a Gaussian kernel centered on the continuous true score, which is then sampled at each fixed bin location.

Given a continuous score $s$, we generate a soft target vector $\boldsymbol{y}$ over $K=20$ bins, where the probability for the k-th bin with center $c_k$ is proportional to
\begin{equation}
y_k \propto \exp\left(-\frac{(s - c_{k})^2}{2\sigma^2}\right).
\end{equation}
The resulting vector is then normalized to sum to one, forming a valid probability distribution. This approach ensures that the probability mass is highest at the bin closest to the true score and decays smoothly for neighboring bins, hence encoding the ordinal information.

% As an alternative to our proposed Gaussian-softened label approach, we also explored a more explicit ordinal regression framework based on the CORAL method \cite{coral}. This method recasts the problem of predicting a score from $K$ ranks (e.g., $K=5$ for MOS) into learning $K-1$ cumulative probabilities. In principle, the model is trained to predict the cumulative probabilities $P(Y > k)$ for each rank $k \in \{1, ..., 5\}$, where $Y$ is the true rank. The model architecture remains identical to that described in Section \ref{subsec:model_architecture}, except for the final prediction head.

As an alternative to our proposed Gaussian label softening approach, we also explored a more explicit ordinal regression framework based on the CORAL method \cite{coral}, which reframes the task as learning $K-1$ cumulative probabilities $P(Y > k)\ \forall\ k \in \{1, ..., 5\}$ (named \textbf{DORA-MOS (CORAL)}). We achieve this simply by replacing the final classification head with $K-1$ sigmoid outputs, and leaving the rest of the network unchanged.

% ======================================================

\begin{table}[t!]
\centering
\caption{Ablation study on the training criterion on the test set.}
\label{tab:ablation_loss_functions}
\resizebox{\columnwidth}{!}{%
\begin{tabular}{@{}lcccc@{}}
\toprule
& \multicolumn{2}{c}{\textbf{Music Impression (MI)}} & \multicolumn{2}{c}{\textbf{Textual Alignment (TA)}} \\
\cmidrule(lr){2-3} \cmidrule(lr){4-5}
\textbf{Training Criterion} & SRCC$\uparrow$ & KTAU$\uparrow$ & SRCC$\uparrow$ & KTAU$\uparrow$ \\
\midrule
L1 Loss & 0.966 & 0.858 & 0.903 & 0.733 \\
CE Loss & 0.973 & 0.876 & 0.940 & 0.793 \\
\textbf{Gaussian} & \textbf{0.991} & \textbf{0.931} & \textbf{0.952} & \textbf{0.825} \\
\bottomrule
\end{tabular}%
}
\end{table}

% ======================================================
\begin{table}[htbp]
\centering
\caption{Test set ablation on temporal modeling and pooling method}
\label{tab:temporal_ablation}
\begin{tabular}{@{}llccc@{}}
\toprule
\textbf{Pooling Method} & \textbf{Metric} & \textbf{Transformer} & \textbf{Mamba} & \textbf{Bi-LSTM} \\
\midrule
\multirow{4}{*}{Mean Pooling} & SRCC MI & 0.973 & 0.984 & 0.972 \\
 & KTAU MI & 0.871 & 0.881 & 0.881 \\ % <-- Add KTAU MI values here
 & SRCC TA & \textbf{0.939} & 0.935 & 0.938 \\
 & KTAU TA & 0.798 & 0.802 & 0.798 \\ % <-- Add KTAU TA values here
\midrule
\multirow{4}{*}{Attention Pooling} & SRCC MI & \textbf{0.991} & 0.975 & 0.964 \\
 & KTAU MI & 0.931 & 0.903 & 0.913 \\ % <-- Add KTAU MI values here
 & SRCC TA & \textbf{0.952} & 0.937 & 0.929 \\
 & KTAU TA & \textbf{0.825} & 0.798 & 0.789 \\ % <-- Add KTAU TA values here
\bottomrule
\end{tabular}
\end{table}

% ======================================================

\section{Results}

We evaluated several architectures and training strategies, focusing on system-level metrics as required by the AudioMOS Challenge. We report on the rank-based metrics, particularly SRCC and Kendall's Tau (KTAU), between the predicted and ground-truth scores. Unless otherwise specified, all reported models use the same underlying cross-attention architecture described in Section \ref{subsec:model_architecture}.

\subsection{Ablation Study on Training Criterion}

We first conducted a controlled ablation study to select the most effective training objective for this task. As described in our data analysis, we used our internal stratified data splits for this experiment to ensure that the model selection was based on generalization to unseen data, yet we report results on the original test set for consistency with AudioMOS Challenge papers. We compared three distinct training criteria: 1) a standard L1 regression loss (as used in the baseline), 2) a naive classification with cross-entropy (CE) loss over 20 discrete bins, and 3) our proposed CE loss with Gaussian-softened labels.

% The results, presented in Table \ref{tab:ablation_loss_functions}, demonstrate that Gaussian softening produces a more stable performance on SRCC across MI and TA scores for both the dev set and test set. Anecdotally, when testing with our internal stratified data splits, we saw that the Gaussian softening generalized the best for the unseen dev set. Here, in  Table \ref{tab:ablation_loss_functions}, we can confirm the same findings by looking at the Test Set performance, where the proposed Gaussian method performs the best in SRCC compared to both L1 and CE loss. This confirms our hypothesis that reframing the problem as classification while explicitly encoding ordinal information provides a more robust training signal. Based on these findings, we selected the Gaussian-softened label criterion for all main experiments.

As shown in Table \ref{tab:ablation_loss_functions}, Gaussian‐softened CE achieves the highest SRCC on both MI and TA for test sets. We got the same findings when testing on the original dev set and trained the models with our stratified data. This confirms that framing MOS prediction as an ordinal classification with smooth label distributions yields a more robust training signal. We therefore adopt this criterion in all subsequent experiments.

\subsection{Ablation on the Temporal Modelling and Pooling}
The challenge requires predicting 30-second-long audio sequences at a 24kHz sample rate, which means efficiently modelling the temporal information is important.
%While CLAP and Muq-Mulan internally manage the alignment of audio and text, producing a single embedding, we show that explicitly modelling the temporal information of 2 separate representation models achieves better SRCC performance. We believe this can be explained due to the TA score being a task-specific score that may need a different, specialized alignment between the audio and text modality for understanding the meaning of generated audio quality and prompt/audio alignment. Nonetheless, using Muq models only reduces the audio sequence by a downsampling rate of 240, which means that for 30-second audio, the sequence dimension of the embedding of Muq is still long. For this reason, we then study which model may be more suitable to learn temporal dependencies of the audio input before feature fusion and the pooling method to use. In particular, as temporal modelling, we analyze a self-attention layer as it has been proven to be suitable for modelling long-term sequential dependencies. However, its computational complexity is $O(N^2)$ so we explored with other variants such as Bi-directional LSTM (BiLSTM) which process sequence at $O(N)$ yet it may not be able to process efficienclty the long audio sequence. Finally, a state-space model, Mamba, is also tested.
Hence, we explore different sequence modelling methods that have different time complexities to see if the transformer component in the temporal modelling block could be replaced. We compare three models: (1) Transformer (self-attention) \cite{transformer}, (2) Mamba \cite{mamba}, (3) Bi-LSTM \cite{blstm}.

% is crucial. We compare three temporal encoders: 
% \begin{itemize}
%     \item \textbf{Self‐Attention}: captures long‐range dependencies but costs $\mathcal{O}(N^2)$.  
%     \item \textbf{Bi‐LSTM}: linear complexity $\mathcal{O}(N)$, more efficient but may struggle with very long sequences. 
%     \item \textbf{Mamba} (State‐Space Model): linear complexity $\mathcal{O}(N)$, designed for scalable, efficient long‐sequence modeling.
% \end{itemize}

For the pooling method, we tested either Average Pooling or Attention Pooling. We hypothesize that Attention Pooling will outperform Mean Pooling. This is because not all segments of a music piece contribute equally to its perceived quality or text alignment. Attention Pooling allows the model to learn to focus on the most salient moments, whereas Mean Pooling treats all time steps as equally important.

The results of the temporal modelling architecture and pooling method can be seen in Table \ref{tab:temporal_ablation}. 
The transformer model with self-attention combined with a learned Attention Pooling performs the best across both ranking metrics on MI and TA. We saw this same behavior when analyzing each method in our inner train dev split, and then we decided to choose that configuration for our system.

Interestingly, the choice of pooling mechanism shows different interactions with the encoder architecture. While Attention Pooling is optimal for the Transformer, Mean Pooling works well for the linear-time models (Mamba and BiLSTM). This suggests that when an encoder builds its context sequentially, a global average across all time steps may be enough.

\begin{table}[t!]
\centering
\caption{System-level performance on the test set measured by SRCC and KTAU. The upper section lists individual models, and the lower section lists ensemble models. The \textbf{\textcolor{red}{Ensemble Ridge}} system was our final submission.}

\label{tab:evaluation_results}
\resizebox{\columnwidth}{!}{%
\begin{tabular}{@{}lcccc@{}}
\toprule
& \multicolumn{2}{c}{\textbf{Music Impression (MI)}} & \multicolumn{2}{c}{\textbf{Textual Alignment (TA)}} \\
\cmidrule(lr){2-3} \cmidrule(lr){4-5}
\textbf{Configuration} & SRCC$\uparrow$ & KTAU$\uparrow$ & SRCC$\uparrow$ & KTAU$\uparrow$ \\
\midrule
CLAP              & 0.818 & 0.623 & 0.724 & 0.532 \\
MuQ-Mulan                      & 0.942 & 0.797 & 0.847 & 0.701 \\
DORA-MOS       & \textbf{0.991} & \textbf{0.931} & 0.952 & \textbf{0.825} \\
DORA-MOS (CORAL)        & 0.975 & 0.885 & 0.920 & 0.766 \\
DORA-MOS (Decoupled)      & 0.983 & 0.908 & \textbf{0.954} & \textbf{0.825} \\
\midrule
\textbf{\textcolor{red}{Ensemble Ridge}} & 0.988 & 0.913 & 0.944 & 0.809 \\
Ensemble LightGBM              & 0.985 & 0.913 & 0.917 & 0.754 \\
\bottomrule
\end{tabular}%
}
\end{table}
\vspace{-1mm}
\subsection{Single System Performance}
%While CLAP and Muq-Mulan internally manage the alignment of audio and text, producing a single embedding, we show that explicitly modelling the temporal information of 2 separate representation models achieves better SRCC performance. We believe this can be explained due to the TA score being a task-specific score that may need a different, specialized alignment between the audio and text modality for understanding the meaning of generated audio quality, as well as prompt/audio alignment. Nonetheless, using Muq models only reduces the audio sequence by a downsampling rate of 240, which means that for 30-second audio, the sequence dimension of the embedding of Muq is still long. For this reason, we then study which model may be more suitable to learn temporal dependencies of the audio input before feature fusion, as well as the pooling method to use. In particular, as temporal modelling, we analyze a self-attention layer as it has been proven to be suitable for modelling long-term sequential dependencies. However, its computational complexity is $O(N^2)$ so we explored with other variants such as Bi-directional LSTM (BiLSTM) which process sequence at $O(N)$ yet it may not be able to process efficienclty the long audio sequence. Finally, a state-space model, Mamba, is also tested.

Given the previous ablations, we selected the architecture with Transformer, Attention Pooling, and Gaussian softening labels (DORA-MOS) as our main model. Table \ref{tab:evaluation_results} reports results on the official dev and test sets. Our top model, \textbf{DORA-MOS}, attains MI SRCC 0.991 (vs. 0.818 baseline) and TA SRCC 0.952 (vs. 0.724 baseline). This result shows the importance of explicitly modelling temporal dependencies from audio and text encoders. To further compare this hypothesis, we test another model that learns audio/text alignment during pre-training to add another comparison to CLAP. In particular, we change CLAP for MuQ-Mulan. It can be seen that while improving the baseline, MuQ-Mulan still considerably lags behind our proposed method, especially on the TA score. This result demonstrates the value of separate audio/text encoders, cross-attention fusion, and Gaussian‐softened labels. Other variants \textbf{CORAL} and \textbf{Decoupled} also outperform the baseline, but the Gaussian‐softened approach remains the best single model.
  
%The performance of our main models on the official challenge Dev and Test sets is detailed in Table \ref{tab:evaluation_results_system}. Our best-performing single model, the \textbf{Cross-Attention-Gaussian}, outperforms the official baseline across all metrics. On the Test set, our model improves the MI SRCC from 0.8176 to 0.9907 and the TA SRCC from 0.7241 to 0.9519. Such results prove the importance of encoding audio and text embeddings separately with their own encoders, combined with a task-specific cross-attention fusion and our proposed training criterion.

% Our other architectural variations, including the \textbf{Cross-Attention-Ordinal Regression} model and the \textbf{Cross-Attention-Decoupled} model, also show strong performance, consistently surpassing the baseline. However, the Gaussian-softened classification approach remained the most effective single model.

\subsection{Stacking Ensemble Strategy}

To further improve prediction accuracy, we employed a two-level stacking ensemble strategy. Stacking involves training a meta-model (Level-1) to aggregate the predictions of multiple base models (Level-0). In Level-0, we combined five variants of our \textbf{DORA-MOS} model (trained with different random seeds), two \textbf{CORAL} models, and two \textbf{Decoupled} models.

% The pool of our Level-0 base models comprised textbf{Cross-Attention Gaussian} model trained with different random seeds (5 variations), as well as \textbf{Cross-Attention-Ordinal Regression} models (2 variations) and two variations of \textbf{Cross-Attention-Decoupled}.

Rather than using only their scalar outputs, we extracted each model's discrete probability distribution over the 20 score bins. For models that produced only a single score, such as the ordinal regression models (CORAL), instead of transforming it to a one-hot Kronecker delta distribution, we applied our Gaussian softening kernel, centered on the predicted value, to generate a smooth 20-bin distribution. The logic for this decision is to avoid the meta-model being biased towards the Kronecker delta predictions. These distributions were concatenated into a high-dimensional feature vector for each audio clip.

% Instead of using only the final scalar predictions from these models, we used their full predicted distributions over the 20 score bins as features for the meta-model. For models that natively produced a distribution, we used it directly. For models that output a scalar score, such as the ordinal regression models, we transformed their prediction into a 20-bin distribution. Rather than treating them as a Dirac distribution, we smooth them out by using the same Gaussian softening kernel centered at the predicted score, as done for the other trained models. We did so to avoid the final meta-model biasing too much towards the Dirac scores. The resulting distributions were then concatenated, forming a high-dimensional feature vector for each audio sample to be used as input for the Level-1 meta-model.

At Level-1, we compared a linear Ridge Regression \cite{hoerl1970ridge} meta-model against a LightGBM \cite{lightgbm} regressor. We simulated generalization by splitting the official development set into 60\% meta‐training and 40\% meta‐validation partitions. Although LightGBM achieved a near‐perfect fit on the meta‐training data (almost 0 Mean Square Error and 0.999 SRCC), Ridge Regression generalized more robustly on the held‐out validation set. Consequently, we selected Ridge Regression for our final submission. Despite this careful stacking procedure, the ensemble yielded slightly lower SRCC on the official test set than our single DORA-MOS model (0.988 vs 0.991 SRCC MI and 0.944 vs 0.952 SRCC in TA).

\subsection{How our system compares with the participating ones}

In the final evaluation, our submitted Ensemble Ridge model was ranked against all participating systems across 16 official metrics, covering utterance and system-level performance for both MI and TA. Our system achieved first place in 13 of 16 metrics, demonstrating robust performance across a comprehensive set of criteria.

Table \ref{tab:system_comparison_no_baseline} details the comparison of the primary system-level rank correlation metrics for all nine teams. Our system obtained the highest scores for MI SRCC (0.988), MI KTAU (0.913), and TA SRCC (0.944). For the Text Alignment KTAU metric, our system secured the second-highest score (0.809). This performance positioned our system as the top performer overall in the challenge.

\begin{table}[t!]
\centering
\caption{Test set rank metric comparison across all challenge systems}
\setlength{\tabcolsep}{4pt}
\label{tab:system_comparison_no_baseline}
\resizebox{\columnwidth}{!}{%
\begin{tabular}{@{}lcccccccccc@{}}
\toprule
\textbf{Metric} & \textbf{T02} & \textbf{T05} & \textbf{T07} & \textbf{Ours} & \textbf{T10} & \textbf{T11} & \textbf{T21} & \textbf{T22} & \textbf{T23} \\
\midrule
SRCC MI & 0.954 & 0.922 & 0.869 & \textbf{0.988} & 0.965 & 0.966 & 0.972 & 0.970 & 0.975 \\
KTAU MI & 0.848 & 0.761 & 0.706 & \textbf{0.913}	& 0.848	& 0.862	& 0.876	& 0.871	& 0.885\\
SRCC TA & 0.941 & 0.839 & 0.807 & \textbf{0.944} & 0.904 & 0.921 & 0.916 & 0.930 & 0.887 \\
KTAU TA & \textbf{0.818}	& 0.652	& 0.615	& 0.809	& 0.767	& 0.776	& 0.763	& 0.786	& 0.721 \\
\bottomrule
\end{tabular}%
}
\end{table}
\section{Conclusions}

In this work, we presented the first-place solution to the AudioMOS 2025 Challenge Track 1. The proposed model shows that explicitly modeling the temporal sequence from separate representation models for both the audio and text modality is important, and by learning a cross-attention fusion, we can significantly improve the modeling of audio/text alignment for a better TA score. While cross-attention fusion is a known technique, our key contribution is an ordinal-aware training criterion. By reframing MOS prediction as a classification task and using Gaussian-softened labels, we successfully embed the ordinal nature of the scores into the learning process. This simple approach significantly improved the SRCC for MI and TA scores. 

As future work, we plan to model annotator disagreement. Namely, instead of predicting a single averaged score, we could adopt a probabilistic framework to predict a distribution that captures annotator disagreement \cite{metaperser}. For example, one could model the output as a Beta distribution to represent the range and variance of human scores \cite{jeremy_beta,beta_mos}. 

Finally, the high scores achieved by our system suggest that performance on the current dataset may be approaching saturation. We think a possible direction for this dataset is to use it for few-shot adaptation. One could evaluate a model's ability to perform few-shot adaptation. The goal would shift from pure prediction to how quickly a model can learn the corpus bias from human MOS scores  \cite{itu2016800,alignet} within this dataset.

% Look at how the model predicts the text alignment and how this changes across systems. It is important to draw insights on this aspect because it seems the TA alignment model learns to generate scores near the distribution mode. With this, you can hack a good score....

% Encoder: [MuQ, MuQ-Mulan, CLAP, MERT] -> maybe not needed to show.... => 4 exps, same setup as cross-attn Gaussian

\newpage
\bibliographystyle{IEEEtran}
\bibliography{mybib}

\end{document}